New Generalized Informatics Framework for Development of Large Scale 'Virtual' Battery Material Databases

Scott R. Broderick [1], Kaito Miyamoto [1,2], and Krishna Rajan [1,*]

[1]Department of Materials Design and Innovation, University at Buffalo, Buffalo, NY 14226, USA

[2] Toyota Research Institute of North America, Toyota Motor North America, Inc., 1555 Woodridge Avenue, Ann Arbor, MI 48105, USA

*krajan3@buffalo.edu

In this paper, we introduce an approach for the prediction of capacity for over 100,000 spinel compounds relevant for battery materials, from which we propose the 20 most promising candidate materials. In the design of batteries, selecting the proper material is difficult because there are so many metrics to consider, including capacity which is a fundamental engineering property. Using reported experimental data as our starting point, we demonstrate how we can build a dataset that provides a guide for the selection of battery materials. Although we focus on capacity of Li based spinel structures for electrode materials relevant for usage in batteries, the methodology developed and demonstrated here can be adapted to other properties, structures, and site occupancies. Further, theoretical capacity is often used as a guideline for material design of battery materials. In this paper, we show how this is insufficient for representing experimental measurements, while our methodology closes this gap and provides an accurate computational representation of experimental data.

I.  Introduction

A challenge in the design of most engineering properties, including capacity, is that no equation exists to link the property with chemistry or structure. This is largely due to the requirements for a sufficiently diverse dataset and the mathematical complexities in linking experimental data with relatively large uncertainty levels associated with it and the complexities associated with changes in site chemistry. While there have been several successes in the machine learning based design of materials [1-6], the examples of accelerated material design specific to batteries have been limited. This is due largely to the requirement of large amounts of data in order to build a robust machine learning model. While density functional theory (DFT) calculations, particularly as provided in the Materials Project database [7-8], have given some guidance on the properties of relevant materials, this still loses the embedded multiscale properties of materials, and is limited to chemistries containing four or fewer elements.

To address these challenges, we propose and demonstrate a machine learning approach to suggest capacity characteristics of battery materials. In order to do this, we create a small training set and then show how our hybrid machine learning approach can create a synthetic database, with the new data set helping to guide the next generation of battery materials. This overcomes the challenges of unbalanced and sparse data by modifying our prior works in defining the data manifold which dictate the property prediction models [1,2,46-51].

For the Li-Me-O system (where Me refers to a metal or multiple metals), if we allow Me to represent three different elements, we have nearly 25,000 possible combinations, and if expanded to four elements, the search space becomes nearly 700,000 possible combinations. These numbers are independent of compositional modifications, which leads to an infinite number of combinations. Additionally, there are several existing and discovered limitations: the amount of experimental data for a sufficiently wide range of chemistries is limited; the application of standard regression approaches does not lead to a meaningful model; and the role of chemistry versus structure/microstructure is undefined. Here, we develop an approach that is applicable to relatively sparse data, while maintaining reasonable robustness, and describe the development of a nearly infinite database for battery materials.

Many reports on different Li-based materials are available and contain data relevant for our analysis. These include layered $LiMeO_2$ materials [9-13], olivine structured materials with $LiMePO_4$ form [14-20], spinel structured materials of $LiMe_2O_4$ [21-26] with much of the focus on $Li_4Mn_5O_{12}$ [27-30], and numerous other possibilities [31-42], although our primary focus in this paper is on the spinel structure. These reports represent our starting knowledge base (Figure 1 defines our initial search space based on the chemistries represented in these reports). While we know that the microstructure of these materials has a large impact on properties [43-44], we are not considering that explicitly here and are focused solely on the chemical implications.

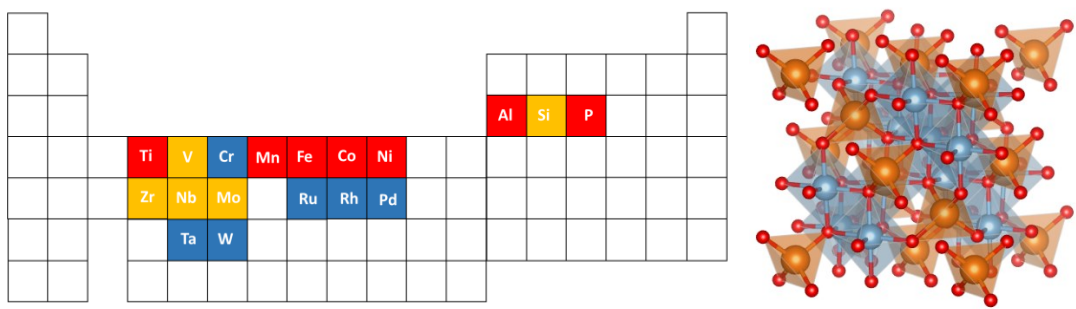

*Figure 1. (Left) Chemical search space for our models, with the elements occupying Me in Li-Me-O, with Me consisting of up to three of the labeled elements (providing over 4,000 possible combinations, independent of composition). Based on availability of training data, we have ranked our confidence in predictions with these elements: red – high confidence, orange – medium confidence, blue – low confidence. (Right) Spinel structure. The elements labeled are used to occupy the blue sites in this figure, with the orange sites and red sites occupied by Li and O, respectively.*

We have extensive work in the development of high-throughput models for the discovery of new compounds across a variety of material platforms [45-51], which we build upon here. In addition to guiding future experiments, our approach adapts with additional data / physics, and thereby is not disconnected with experimental efforts, but rather reduces uncertainty with each iteration, and opens up the chemical search space of battery materials.

II.  Data

We have developed a new hybrid methodology which allows for robust predictions with sparse training data, and which captures the complexity of the physics of the problem through the introduction of non-linearity. This methodology (Figure 2) is developed in a generalized manner so that it can incorporate all types of data. Further, a uniqueness of our approach is that it is not based on composition, but rather the descriptor space; thus, the models are not based on fitting to composition and thereby limiting the search space, but rather based on the underlying physics.

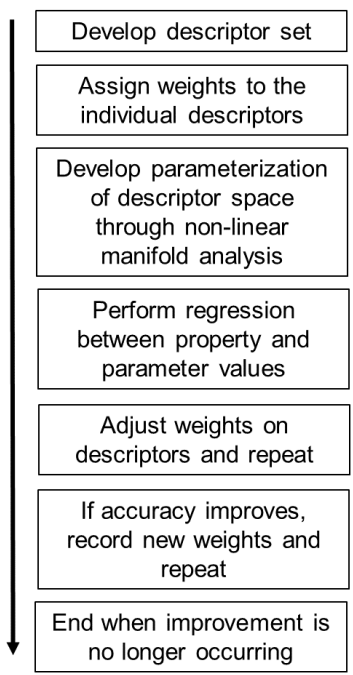

*Figure 2. The developed hybrid methodology for predicting material properties. This approach has been newly developed here, and links multiple techniques, including non-linear manifold with regression techniques. The two aspects of data generation described in this section are the development of descriptor sets and the extraction of property data from literature.*

The descriptors used are based on various characteristics from the periodic table, as described in our prior works [1-2,50]. These encompass various characteristics such as sizes of atoms, electronegativities and single element properties. The additional step then is to represent the material by the descriptors. The scaling of elemental descriptors to represent material properties is justified by the works of Villars, Mooser-Pearson, Pettifor, and Hume-Rothery [52-54].

The descriptor set is based on the understanding that chemistry is site specific, so we are confining elements to the sites which they have been experimentally shown to occupy. In this way, we are building in a physics-based constraint to increase the likelihood that we are considering material chemistries that can exist. Here the descriptors are defined as the sum of the descriptor values for the cation components (such that the input descriptor for radius for $LiTi_2O_4$ would be ($r_{Li}$ x 1/3) + ($r_{Ti}$ x 2/3). The input dataspace represents the characteristics of the materials rather than the chemistry, and thus the objective is to identify why materials have their behavior, as opposed to a solely statistical exercise. For the measurements of capacity (from the previously mentioned references), there is some inconsistencies in values. This is expected as there are numerous changes occurring beyond just chemistry, as well as the expected uncertainty in experimental measurements. To limit any bias in the analysis, we have taken the average value of capacity for a given chemistry. The input descriptor space is shown in Figure 3, with the descriptors listed in the Supplementary Material.

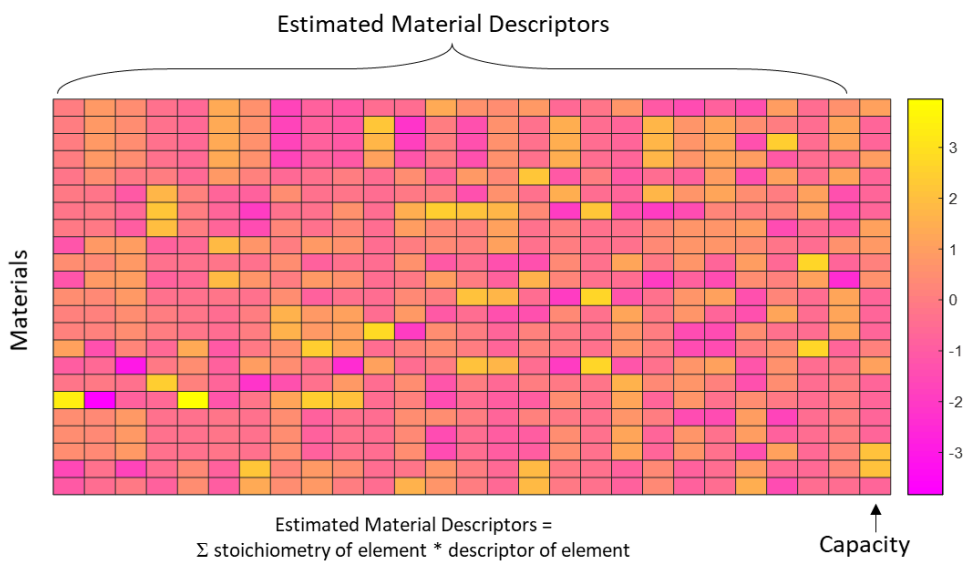

*Figure 3. Conversion of descriptor space into parameterized space. The descriptors incorporate multiple aspects of materials. In this figure, we see reasonable distribution, so that the models are not overly sensitive to any outlier measurements. This represents the input into our data analyses.*

III. Methodology

A quantitative model for capacity was developed based on a hybrid connection between regressions and IsoMap. In this way, we combine a regression approach which accounts for inter-correlations in the data, while adding in the consideration of a non-linear data manifold. For property prediction, we used the descriptor set and input it into a graph analysis. This provides a set of parameters, which capture the non-linear relationships in the descriptors. The reason for doing this is to reduce the dimensionality of the input without losing information. That is, we use the graph theory for dimensionality reduction, and thereby can perform a regression with limited risk of over-fitting the model. The reason is that we want to create a parameter space which can be used for all systems – that is, we develop a model based on graph theory parameters, and we need these parameters for all systems of interest. As more systems are included, this step would need to be repeated to create the parameters for the new systems as well as capture any other changes due to updated information. This entire process requires multiple steps, each of which have their own complexities. We expand on the steps as follows.

*Step 1: Convert design space to a graph network* (Figure 4). The input data described spans represents an n-dimensional space. Within this space, inter-correlations exist and the relationships are hard to define. Further, a standard regression approach or even principal component regression approach is insufficient for accurate modeling (as discussed in the results section). To account for these challenges, we first develop a non-linear parameterization of the data through non-linear manifold learning, and specifically the IsoMap algorithm [46-47,50,57]. This approach generates a graph connecting data points on a high dimensional space to their nearest neighbors, mapped out in the high dimensional space, and then fit to a low dimensional manifold. Through dimensionality reduction the manifold unravels into two or three dimensions allowing it to be visualized. We develop a set of parameters for each set of conditions, although in this case the parameters are based on a non-linear combination of descriptors. Initially, the weights on each descriptor are set to unity (ie. the descriptor values are not scaled or manipulated, beyond standardization). As discussed in the next step, to improve the prediction, weights are applied on each descriptor to have varying levels of their impact on the model. These weights are initially randomly selected and are then updated to screen the potential descriptor space. This process is repeated until the combination of weights which leads to the best model are identified. This added step of defining weights is akin to the training process in genetic algorithm, but simplified in this new approach defined here to account for the small data size.

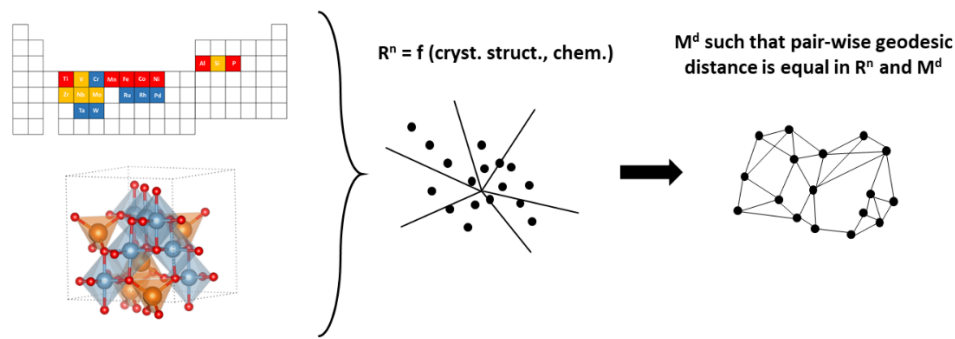

*Figure 4. The conversion from our training data onto a non-linear manifold and a graph network. Using the search space and constraints defined in Figure 1, we developed an input data set for training our models. These inputs provide a high-dimensional description, which we then convert into a manifold learning space which provides the inputs for our prediction.*

*Step 2: Test that the graph network captures the property* (Figure 5). The regression operates by defining the series of non-linear manifold parameters that are then used to predict the capacity through a multi-linear regression. Cross validation is employed to test the robustness of the model where a comparison is made between accuracy of training versus test data. To do this, we performed multiple iterations removing different 1/3 groupings of data prior to the regression and using these for testing the model and to ensure that the data is not over-fit. This step also defines the accuracy for the selection of descriptor weights. Steps 1 and 2 are continually repeated with new weightings until there is no further improvement in the test accuracy.

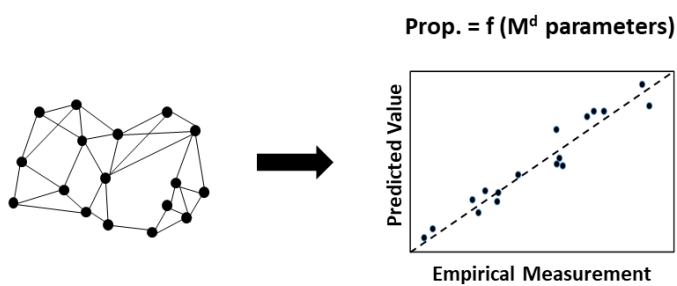

*Figure 5. The graph network built from the prior step provides a parameterization of the data. These parameters capture the complexity of the correlations in the data and account for the non-linearity in these relationships. These parameters are then used to develop a predictive model. The quality of the model is determined by the relationship of the predicted and training data, with $R^2$ value closer to unity desired.*

*Step 3: Develop a 'virtual' dataset and build new manifold* (Figure 6). From the prior steps, the descriptor set was developed and the approach for linking the site chemistry with the property has been tested. The main objective though is to predict the property (capacity here) for materials which we do not have data for. The descriptors input into the analysis are all based on the chemistry and can be generated for any combination of materials. That is, this process does not require any experimental data or any additional data generation. The use of the experimental data is only in training the model between the parameterization and the property, and then for validating the model. The number of 'virtual' materials (ie. materials that have not been made or tested) that can be explored is nearly infinite. We here have confined the substitution of elements to the B-site of the spinel structure to only incorporate three different elements. The descriptors are then automatically generated based on the chemistry, and then the process of developing the network and parameterizing the data with a non-linear manifold are repeated.

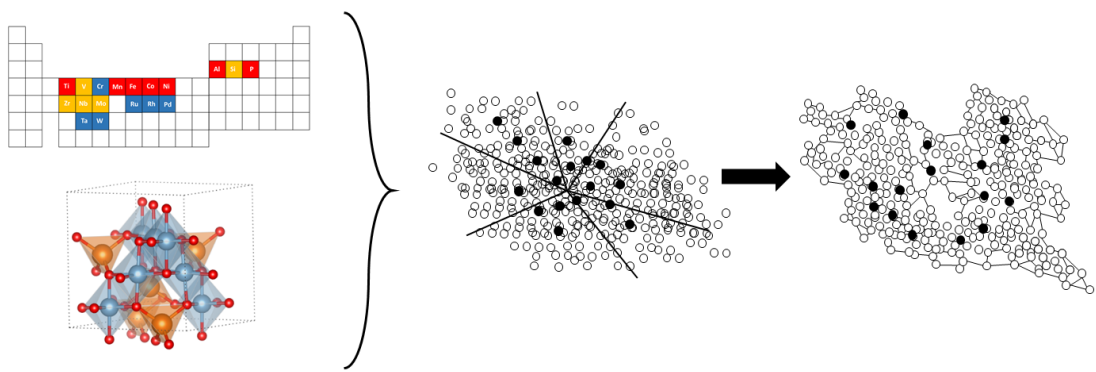

*Figure 6. Using the same design space, we generate a new series of 'virtual' materials, which are defined by the same descriptors as in the first two steps. From this descriptors space, a new graph network and material parameterization are generated. In this figure, the filled black circles are the known data used in the prior steps, while the open circles represent new materials that have not been previously tested.*

*Step 4: Build an updated model for prediction of 'virtual' materials* (Figure 7). The process for developing a predictive model is the same as in step 2. That is, the same chemistries used to define the relationship between the network parameterization and the capacity are used. The parameter values have changed between the two steps because the new network contains many more points. The number of chemistries used in building the regression and the split between training and test data is the same as before. This model can then be applied to all of the 'virtual' chemistries and the output is a predicted capacity value.

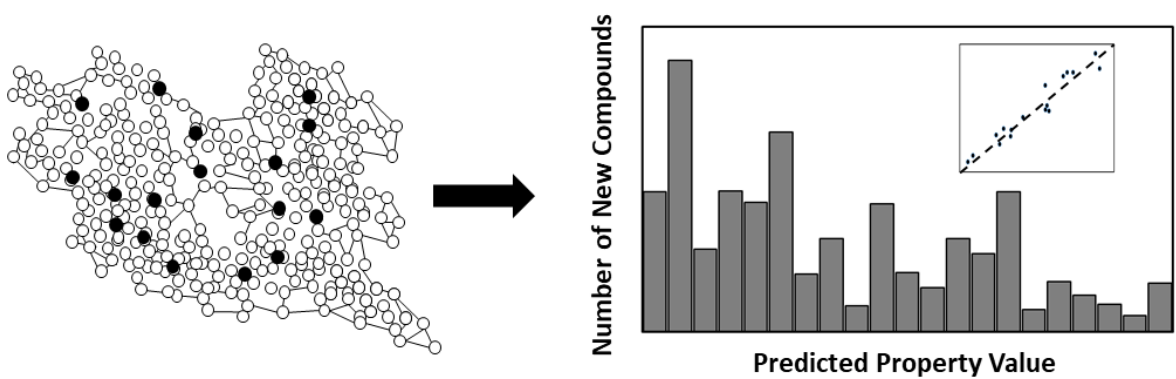

*Figure 7. Repeat of the regression approach, but using the new parameterization set which encompasses the 'virtual' chemistries. The filled black circles are used for developing the predictive model. After ensuring a nearly linear relationship between the predicted and the measured (inset of right figure), the model is applied to all of the materials represented by open circles. The values for all of these represent a new synthetic material database, with a distribution in property values following that as shown. Those with the highest value can then be identified, providing a significant screening of the massive search space.*

*Step 5: Using model, develop a synthetic database for material properties* (Figure 7). The model is applied to all of the 'virtual' chemistries and the predicted properties are tabulated along with the chemistry and the descriptors. In this way, a new database with a massive increase in number of reported property values is developed. In the results section of this paper, we describe the specifics for developing this database for capacity of Li-based spinel compounds, as well as the screening of the descriptors to select and propose the target materials for further examination.

IV. Results

Compressing all of the information contained in the data into a design map, allows for high throughput modeling which to date has not been possible. To provide an overview of the correlations (accounting for co-linearity of relationships) between chemical descriptors and capacity, principal component analysis was applied and the correlations between descriptors and capacity were defined. The average atomic radius, scaled electronegativity, atomic weight, elemental electrical resistivity, and average pseudopotential radius were identified as the most important design considerations.

A standard objective in materials design is to identify the minimum amount of information needed to design a material. Beyond the obvious implications in data generation, it also serves to identify the inputs into the model. The benefit of minimizing the dimensionality in the data is to reduce the risk of over-fitting the data. Based on these input descriptors and using the cross-validation described in the prior section, and with two latent variables (ie. PCs) employed, the general trend in the data is captured but the overall accuracy ($R^2$ value of 59.7%) of the model is insufficient for generating further predictions. The reasoning for this is due to either an insufficient number or diversity of descriptors, or else the physics governing the system is unable to be captured through the linear mathematics. For this reason, the non-linear parameterization approach was applied to develop a new set of parameters. It is anticipated that the general correlations are correct, but this defines the problem that we have, in that reducing the dimensionality is not sufficient for capturing the entire complexity of information, although we also have the challenge of not over-fitting the model. This challenge defines our need to develop an alternative approach.

Using the entire descriptor set, the non-linear parameterization was used to generate a set of descriptors, which capture non-linear relationships while maximizing variance captured and therefore provides a set capturing the complexity of the information of the input space while having few parameters and thereby avoiding over-fitting of the data. The overview of the parameterization is shown in Figure 8, where each point (node) represents a material chemistry. In this 2D mapping, the first two parameters correspond with the axes, with four total parameters included as that captures over 90% of the variance in the data. Further, five nearest neighbor connections were used in defining the parameters, as that was the maximum number of descriptors prior to short-circuiting the network. As an example of interpretation, the node for $LiTi_2O_4$ labeled and the four parameter values shown. These reduced sets provide a representation of the overall descriptor set, capturing their complexity, and serves as the input parameters for the prediction of the corresponding capacity. A set of four parameters is captured for each material (node).

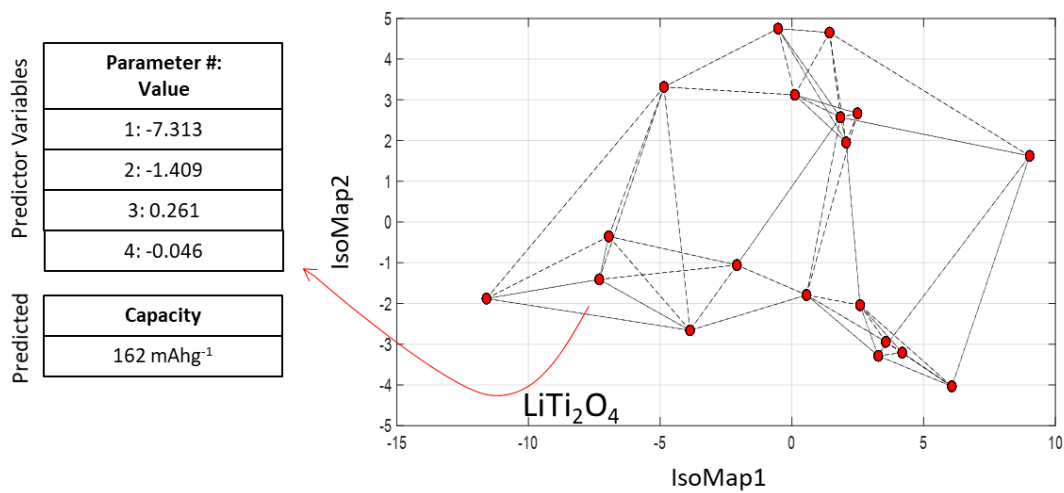

*Figure 8. Non-linear parameterization of the data. Each node represents a different material. The reduced parameter space provides a set of descriptors which capture the complexity of the physics while providing an input which minimizes the over-fitting of data.*

The prediction process was then repeated using these parameters instead of the identified descriptors, with the result shown in Figure 9. To assess the role of chemistry versus structure on the capacity, the process was repeated using only the spinel systems in the training data and predicting properties for only spinel systems, resulting in an accuracy improved to a level sufficient for design ($R^2 = 89.3\%$) and within the expected level of uncertainty associated with experimental measurements. Additionally, we have different levels of confidence in the predictions based on the amount of training data, as defined in Figure 1. Therefore, we have developed a model for predicting the entire search space for spinel chemistries for applications as a battery material with an associated confidence metric.

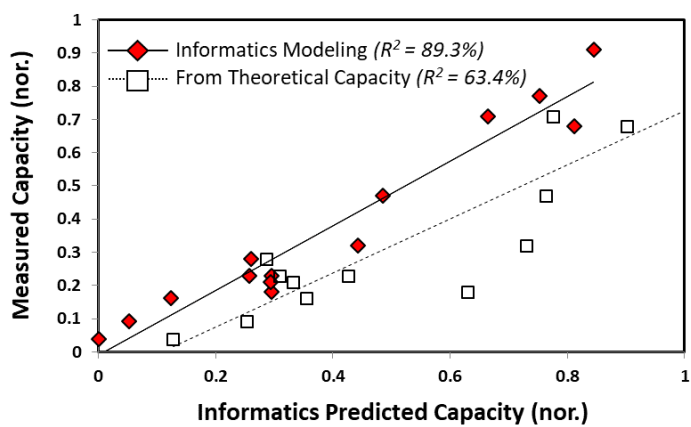

*Figure 9. Informatics prediction of capacity for spinel compounds. The theoretical capacity values were taken from the Materials Project. We have developed an approach and model which allows us to explore all of the spinel structure compounds with a clear level of information added over the use of theoretical capacity.*

This result demonstrates the merit of our machine learning model over using theoretical capacity. That is, it is difficult to predict experimental capacity from theoretical capacity, but the approach laid out and used here can predict experimental capacity using only available data. The expansion to unknown systems uses the following steps: (i) develop a randomized set of material compositions which can exist within the search space (here we do that for 125,000 randomly selected compositions within spinel stoichiometry and using only elements in our training data, although the number could be expanded to an infinite number, although we are sufficiently sampling the feasible search space with some confidence); (ii) develop the descriptor set for those compositions, following the approach laid out in this paper; (iii) develop the graph network / non-linear parameterization and extract the parameters for the appropriate number of dimensions for each of these compositions; (iv) for those compositions which we have the corresponding property (eg. capacity), develop a regression model between the non-linear parameters and the property; (v) apply this model to the large set of parameters for the computationally designed compositions. Table 1 compares the results for the different steps in our computation design methodology. This demonstrates that the training accuracy of the model is maintained even after including the vast

'virtual' materials, further implying that the underlying physics we are modeling is represented in our input descriptor set.

Table 1. Comparison of modeling approaches. For the case of the multiple linear regression (MLR) and principal component regression (PCR) approaches, the input data was our reduced descriptor space. In both cases, the accuracy was insufficient for guiding future design. However, when employing the developed modeling approach outlined in Figure 2, we improved accuracy significantly. In the case of 'only training data', the graph network parameterization used only those systems which we have capacity data for. In the case of 'with 'virtual' systems', the network included also our newly developed material space. This demonstrates that we did not experience significant decrease in accuracy with the addition of the 'virtual' systems.

| | |
|---|---|
| MLR Prediction | 47.1% |
| PCR Prediction | 59.7% |
| **Graph Based Prediction (only training data)** | **89.3%** |
| **Graph Based Prediction (with 'virtual' systems)** | **87.6%** |

This methodology has resulted in capacity values for 125,000 spinel chemistries. In order to consider the data, we are able to screen it for those chemistries which fall in a target range, purely through a data retrieval. As an example, Table 2 lists the 20 chemistries which are predicted to have the highest capacity, considering only the material. This search could be further refined to limit specific elements (for considerations such as cost, difficulty in working with some elements, existing knowledge on microstructural considerations, and so on), and therefore it is designed to be easily interfaced with existing domain knowledge. This highlights further the importance of the approach we have laid out in this paper. That is, the amount of data provided in the literature is rather small, and thus the material search space has not been sufficiently represented in the knowledge base. Thus, approaches to address this issue are necessary. We have assessed the confidence of our predictions and put a relative confidence categorization. This was done through Rough Set Theory [1,60], with the results providing a confidence level for the prediction.

*Table 2. The systems predicted to have the highest capacity for spinel compounds. These reported are limited to spinel compounds of the form LiMe$_2$O$_4$. We have thus reduced the infinite search space to a targeted 20 compounds for experimental testing. The confidence level for each predicted compound is provided from the RST analysis (Section S3).*

| | | | |
|---|---|---|---|
| LiRuPdO$_4$ | High | LiTiCo$_{0.5}$M$_{0.5}$O$_4$ | Low |
| LiRhPdO$_4$ | High | LiRu$_{0.5}$Co$_{0.5}$NiO$_4$ | Low |
| LiMn$_{0.5}$Mo$_{0.5}$RuO$_4$ | High | LiAl$_{0.5}$Mn$_{0.5}$CoO$_4$ | Low |
| LiNi$_{0.5}$Rh$_{0.5}$PdO$_4$ | High | LiTaWO$_4$ | Low |
| LiMn$_{0.5}$Nb$_{0.5}$RuO$_4$ | Medium | LiCo$_{0.5}$RuW$_{0.5}$O$_4$ | Low |
| LiMn$_{0.5}$Mo$_{0.5}$RhO$_4$ | Medium | LiTiCoO$_4$ | Low |
| LiAl$_{0.5}$MnCo$_{0.5}$O$_4$ | Medium | LiMn$_{1.5}$Co$_{0.5}$O$_4$ | Low |
| LiMnNi$_{0.5}$Co$_{0.5}$O$_4$ | Medium | LiMnCo$_{0.5}$Ti$_{0.5}$O$_4$ | Low |
| LiMnV$_{0.5}$Co$_{0.5}$O$_4$ | Medium | LiMn$_{0.5}$Co$_{0.5}$TiO$_4$ | Low |
| LiCo$_{0.5}$V$_{0.5}$NiO$_4$ | Low | LiCo$_{0.5}$Ti$_{0.5}$NiO$_4$ | Low |

V. Discussion

The framework described here and which we have demonstrated for spinel compounds, allows for rapid exploration of new materials for use in battery applications. Of note, are some of the following conclusions from this. The methodology is able to incorporate a larger range of descriptors. For example, the incorporation of processing can be done through the inclusion of additional columns corresponding to processing. In the case of structural descriptors, the definition of descriptors is more challenging due to an inconsistency in the atoms and geometric correlations across structures. A future paper will discuss the structural descriptors which allow us to have a more generalized model, within this framework.

The model was sufficiently accurate to be applied for developing a database on spinel systems. As discussed in the introduction, we have 25,000 possible chemistries for LiMe$_2$O$_4$ if Me represents multiple sites. We further vary the compositions of these Me sites to include composition of 0, 0.25, 0.5, 0.75, or 1, and therefore we have generated a dataset of 125,000 different spinel compounds with the corresponding predicted capacity. Compared to the reports in literature or the vast information available from the Materials Project, this represents a significant acceleration of information (Figure 10). Note, that this gap will significantly expand with the inclusion of other structures beyond spinel compounds.

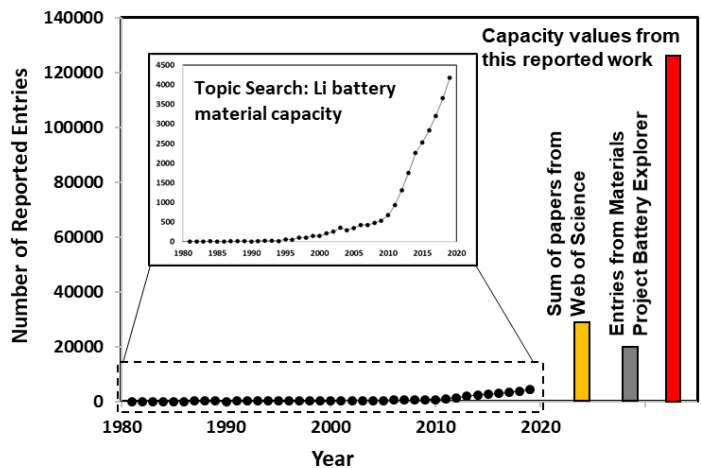

*Figure 10. Acceleration of property values through the data-driven approach described here. The number of papers corresponding with a search of 'Li battery material capacity' in Web of Science shows a notable acceleration in the last decade, highlighting the importance of this topic. Summing the number of papers, as well as the number of materials from the battery explorer (both intercalation and conversion compounds), provides a large starting data space. However, based on only spinel compounds, the number of capacity values reported has been significantly accelerated.*

## VI. Conclusions

This work described a new approach for the development of large-scale databases for new battery materials. A new methodology based on a non-linear manifold parameterization was used to develop high throughput models for capacity. Through these models, we define that the impact of structure versus chemistry in performance while accelerating material property values well beyond what was previously available in all prior measurements combined. The models discussed do not account for the feasibility of these materials to exist, but at the same time do guide the selection of experiments, with the framework developed to adapt to new data, including processing and structure information, as well as other relevant properties. Future work will expand upon the 20 most promising systems outlined here, by considering other structures, additional properties, and the role of processing. However, this provides an important framework to build upon, and which now allows us for the first time to consider all of these considerations simultaneously.

Acknowledgements

We gratefully acknowledge the support of Toyota Motor North America, Inc. SB and KR acknowledge support from the National Science Foundation (NSF) under Grant No. 1640867. KR acknowledges the Erich Bloch Endowed Chair at the University at Buffalo.

# References

1. A. Dasgupta, S.R. Broderick, C. Mack, B.U. Kota, R. Subramanian, S. Setlur, V. Govindaraju, K. Rajan. "Probabilistic Assessment of Glass Forming Ability Rules for Metallic Glasses Aided by Automated Analysis of Phase Diagrams." Scientific Reports, 9, 357 (2019)
2. P.V. Balachandran, S.R. Broderick, K. Rajan. "Identifying the "Inorganic Gene" for High Temperature Piezoelectric Perovskites through Statistical Learning." Proceedings of the Royal Society A. 467, 2271 (2011)
3. C.E. Mohn, W. Kob, Comp. Mat. Sci., 45, 111 (2009).
4. G.H. Jóhannesson, T. Bligaard, A.V. Ruban H.L. Skriver, K.W. Jacobsen, J.K. Nørskov, Phys. Rev. Letts., 88, 255506 (2002).
5. Zakutayev, A., Wunder, N., Schwarting, M. et al. An open experimental database for exploring inorganic materials. Sci Data 5, 180053 (2018).
6. S.V. Dudiy, A. Zunger, Phys. Rev. Lett., 97, 046401 (2006).
7. C.C. Fischer, K.J. Tibbetts, D. Morgan, G. Ceder, Nat. Mater., 5, 641 (2006).
8. S. Adams and R. P. Rao: High power Li ion battery materials by computational design; Phys. Status Solidi A 208, 1746–1753 (2011).
9. Whittingham, M. S., 2004. Lithium batteries and cathode materials. American Chemical Society, 104, pp.4271–4301.
10. Jung, D.S. et al., 2014. Recent progress in electrode materials produced by spray pyrolysis for next-generation lithium ion batteries. Advanced Powder Technology, 25, pp.18–31.
11. Xu, B. et al., 2012. Recent progress in cathode materials research for advanced lithium ion batteries. Materials Science and Engineering R, 73, pp.51–65.
12. Hu, M., Pang, X.L. & Zhou, Z., 2013. Recent progress in high-voltage lithium ion batteries. Journal of Power Sources, 237, pp.229–242
13. Doeff, M. M., 2013. Battery Cathodes. Spring Science+ Business Media,NewYork
14. Nakamura, T. et al., 2007. Electrochemical study on $Mn^{2+}$-substitution in $LiFePO_4$olivine compound. Journal of Power Sources, 174(2), pp.435–441.
15. Dong, Y.Z., Zhao, Y.M. & Duan, H., 2011. Crystal structure and lithium electrochemicalextraction properties of olivine type $LiFePO_4$. Materials Chemistry and Physics, 129(3), pp.756–760.
16. Kamon-in, O. et al., 2013. An insight into crystal, electronic, and local structures of lithium iron silicate ($Li_2FeSiO_4$) materials upon lithium extraction. Physica B: Condensed Matter, 416, pp.69–75.
17. Zhang, W.-J., 2011. Structure and performance of $LiFePO_4$cathode materials: A review. Journal of Power Sources, 196(6), pp.2962–2970.
18. Fey, G. T. et al., 2009. Electrochemical properties of $LiFePO_4$prepared via ball-milling. Journal of Power Sources, 189(2009), pp.169–178.
19. Fey, G. T. & Lu, T., 2008. Morphological characterization of $LiFePO_4$/C composite cathode materials synthesized via a carboxylic acid route. Journal of Power Sources, 178(2008), pp.807–814.
20. Wang, Y. et al., 2011. Morphology control and electrochemical properties of nanosize $LiFePO_4$cathode material synthesized by co-precipitation combined with in situ polymerization. Journal of Alloys and Compounds, 509(3), pp.1040–1044
21. Xu, B. et al., 2012. Recent progress in cathode materials research for advanced lithium ion batteries. Materials Science and Engineering R, 73, pp.51–65.
22. Chemelewski, K.R. & Manthiram, A., 2013. Origin of site disorder and oxygen nonstoichiometry in $LiMn_{1.5}Ni_{0.5-x}M_xO_4$(M = Cu and Zn) cathodes with divalent dopant ions. Journal of Physical Chemistry C, 117, pp.12465–12471.
23. Zeng, Y.-P. etal., 2014. Effect of cationic and anionic substitutions on the electrochemical properties of $LiNi_{0.5}Mn_{1.5}O_4$spinel cathode materials. Electrochimica Acta.

24. Park, S.H. et al., 2007. Comparative study of different crystallographic structure of LiNi0.5Mn1.5O4−δ cathodes with wide operation voltage (2.0–5.0V). Electrochimica Acta, 52(25), pp.7226–7230.
25. Zheng, J. et al., 2012. Enhanced Li+ ion transport in LiNi0.5Mn1.5O4 through control of site disorder. Phys. Chem. Chem. Phys., 2012,14, 13515–13521.
26. Xiao, J. et al., 2012. High-performance LiNi0.5Mn1.5O4 spinel controlled by Mn3+ concentration and site disorder. Advance Materials, 2012, 24, 2109–2116.
27. Thackeray, M.M. et al., 1987. Spinel electrodes for lithium batteries —A review. Journal of Power Sources, 21(1), pp.1–8
28. Takada, T. et al., 1997. Novel synthesis process and structure refinements of Li4Mn5O12 for rechargeable lithium batteries. Journal of Power Sources, 68(2), pp.613–617.
29. Cao, J. et al., 2013. Electrochemical properties of 0.5Li2MnO3·0.5Li4Mn5O12 nanotubes prepared by a self-templating method. Electrochimica Acta, 111, pp.447–454.
30. Choi, W. & Manthiram, A.,2007. Influence of fluorine substitution on the electrochemical performance of 3V spinel Li4Mn5O12−ηFη cathodes. Solid State Ionics, 178(27-28), pp.1541–1545.
31. Harishchandra Singh, Mehmet Topsakal, Klaus Attenkofer, Tamar Wolf, Michal Leskes, Yandong Duan, Feng Wang, John Vinson, Deyu Lu, and Anatoly I. Frenkel Phys. Rev. Materials 2, 125403 – Published 20 December 2018
32. Arora, P. & White, R. E. Capacity Fade Mechanisms and Side Reactions in Lithium-Ion Batteries. J. Electrochem. Soc.145,3647–3667 (1998)
33. Aoshima, T., Okahara, K., Kiyohara, C. & Shizuka, K. Mechanisms of manganese spinels dissolution and capacity fade at high temperature. J. Power Sources98,378–381 (2001).
34. Jang, D. H., Shin, Y. J. & Oh, S. M. Dissolution of Spinel Oxides and Capacity Losses in 4 V Li/LixMn2O4Cells. J. Electrochem. Soc.143,2204 (1996
35. Tsunekawa, H. et al.Capacity Fading of Graphite Electrodes Due to the Deposition of Manganese Ions on Them in Li-Ion Batteries. J. Electrochem. Soc.149,A1326 (2002).
36. Zhan, C. et al.Mn(II) deposition on anodes and its effects on capacity fade in spinel lithium manganate-carbon systems. Nat. Commun.4,2437 (2013).
37. Journal of The Electrochemical Society, 2020 167 040504 ; Nicolas Gauthier,1,2 Cécile Courrèges Julien Demeaux, Cécile Tessier, and Hervé Martinez
38. Materials Today; Volume 18, Issue 5, June 2015, Pages 252-264; NaokiNitta[13] FeixiangWu[123] Jung TaeLee[13] GlebYushin[1]
39. J. Am. Chem. Soc. 2013, 135, 4, 1167–1176 ; John B. Goodenough and Kyu-Sung Park
40. RSC Adv., 2018,8, 18597-18603; Shouzhong Yi,ab Bo Wang, ORCID logo a Ziang Chen,a Rui Wang ORCID logo *a and Dianlong Wang ORCID logo *a
41. Alternative Energy Technologies: Opportunities and Markets ; By Robert N. Castellano ; Archives contemporaines, 2012
42. Liu, X, Li, K and Li, X (2018) The Electrochemical Performance and Applications of Several Popular Lithium-ion Batteries for Electric Vehicles - A Review. In: Communications in Computer and Information Science (CCIS). IMIOT and ICSEE 2018, 21-23 Sep 2018, Chongqing China. Springer Verlag . ISBN 978-981-13-2381-2
43. Nano-Micro Letters volume 12, Article number: 30 (2020 ; Binitha Gangaja, Shantikumar Nair & Dhamodaran Santhanagopalan
44. Journal of Sol-Gel Science and Technology volume 89, pages225–233(2019) Mateusz Odziomek, Frederic Chaput, Frederic Lerouge, Anna Rutkowska, Konrad Świerczek, Dany Carlier, Maciej Sitarz & Stephane Parola
45. A. Dasgupta, Y. Gao, S. Broderick, E.B. Pitman, K. Rajan. "Machine Learning Aided Identification of Single Atom Alloy Catalysts." Journal of Physical Chemistry C, 124, 14158-14166 (2020)

46. A. Mullis, S. Broderick, A. Binnebose, N. Peroutka-Bigus, B. Bellaire, K. Rajan, B. Narasimhan. "Data Analytics Approach for Rational Design of Nanomedicines with Programmable Drug Release." Molecular Pharmaceutics, 16, 1917-1928 (2019)
47. X. Zhen, T. Zhang, S. Broderick, K. Rajan. "Correlative Analysis of Metal Organic Framework Structures through Manifold Learning of Hirshfeld Surfaces." Molecular Systems Design & Engineering. 3, 826-838 (2018)
48. M. Ashton, R.G. Hennig, S.R. Broderick, K. Rajan, S. Sinnott. "Computational Discovery of Stable $M_2AX$ Phases." Physical Review B. 94, 054116 (2016)
49. S.R. Broderick, G.R. Santhanam, K. Rajan. "Harnessing the Big Data Paradigm for ICME: Shifting from Materials Selection to Materials Enabled Design." JOM. 68, 2109-2115 (2016)
50. S. Srinivasan, S.R. Broderick, R. Zhang, A. Mishra, S.B. Sinnott, S.K. Saxena, J.M. LeBeau, K. Rajan. "Mapping Chemical Selection Pathways for Designing Multicomponent Alloys: An Informatics Framework for Materials Design." Scientific Reports. 5, 17960 (2015)
51. S. Broderick, K. Rajan. "Informatics Derived Materials Databases for Multifunctional Properties." Science and Technology of Advanced Materials. 16, 013501 (2015)
52. P. Villars, J. Less Comm. Metals 99, 33 (1984).
53. P. Villars, J. Less Comm. Metals 92, 215 (1983).
54. A.R. Miedema, J. Less Comm. Metals 32, 117 (1973).
55. Ericksson L., Johansson E., Kettaneh-Wold N.& Wold S.. 2001Multi- and megavariate data analysis:principles, applicationsUmea, SwedenUmetricsAb
56. Ringnér M.. 2008What is principal component analysis. Nat. Biotechnol 26.
57. J. Tenenbaum, V. de Silva, J. Langford, Science, 290, 2319 (2000).
58. Wold, S.; Sjöström, M.; Eriksson, L. PLS-Regression: A Basic Tool of Chemometrics. In Chemometrics and Intelligent Laboratory Systems; Elsevier, 2001; Vol. 58, pp 109–130.
59. Nguyen, D. V.; Rocke, D. M. Tumor Classification by Partial Least Squares Using Microarray Gene Expression Data. Bioinformatics 2002, 18 (1), 39–50.
60. S. Ganguly, C.S. Kong, S.R. Broderick, K. Rajan. "Informatics Based Uncertainty Quantification in the Design of Inorganic Scintillators." Materials and Manufacturing Processes, 28, 726-732 (2013)

Supplementary Material :

S1. Mathematical Background

The machine learning approach encompasses multiple stages: the development of a relevant descriptor set, the parameterization of data to avoid over-fitting of the model while maintaining the governing physics, and the development of a high-throughput quantitative structure-property relationship (QSAR). In both cases, the properties were not input into the data parameterization, so as to allow for prediction of unknown or 'virtual' materials.

The parameterization of the data was done following a non-linear manifold learning approach, and namely the IsoMap algorithm [1-4]. This approach generates a graph connecting data points on a high dimensional space to their nearest neighbors, mapped out in the high dimensional space, and then fit to a low dimensional manifold. The objective of the Isomap algorithm is to map the distribution of elements in the high dimensional space, represented by the set of data points $\{x_i\} \in R^n$, onto a convex nonlinear manifold $M^d$ of lower dimension $d < n$ and through dimensionality reduction, obtain a two or three dimensional embedding of the elements into a weighted graph. The mapping is carried out such that the geodesic distances between the elements in the higher dimensional manifold is preserved when it is mapped onto the lower dimensional graph, so that the edges of the graph are weighted in their length according to the original geodesic distances. The dissimilarity between alloying elements, which themselves form the vertices of this graph, are captured by these distances between them along the edges that connect them to their nearest neighbors. This mapping can be described in set theory as: $x_i \rightarrow y_i \mid y_i \in M^d$, $d<n$, s.t. $\forall (i,j): |x_i - x_j|_\beta = |y_i - y_j|_\beta$ where $\beta$ is a norm, representative of the pairwise geodesic distances $d_{ij}$ between any two elements $'i'$ and $'j'$, which is the curvilinear distance along the manifold in $M^d$.

In order to construct the initial graph in $R^n$, we used K nearest-neighbors (KNN), which graphs each point connected by an edge to its $'k'$ nearest neighbors alone ∋ $d_{ij} = \infty$, $\forall |i-j| > k$. Among semi-supervised learning KNN has been found to perform well compared to other graphs [5] and was, therefore, employed in the present work. In this work the choice of $k$ was optimized by statistically determining the smallest value that could minimize the residual variance $\mid d_M - d_G \mid$, while providing the maximum number of alternative paths. This ensures that the resulting graph is neither over-connected, leading to loss of pairwise geodesic distances, nor are critical neighbors disconnected. For each data point, we also compute the ratio of the distance to its closest and farthest neighbor. The ratios are then averaged over all data points to calculate a scale-invariant, global parameter, $\Delta$, [6] to estimate the measure of uncertainity introduced by sparsity in high dimensional spaces, given that the data points must have sufficient density on the manifold [7]. $\Delta$ can range between zero and one and a small value indicates a healthy variance in pairwise distances.

For property prediction, we used the descriptor set and input it into a graph analysis. This provides a set of parameters, which capture the non-linear relationships in the descriptors. The reason for doing this is to reduce the dimensionality of the input without losing information. That is, we use the graph theory for dimensionality reduction, and thereby can perform a regression with limited risk of over-fitting the model. The reason is that we want to create a parameter space which can be used for all systems – that is, we develop a model based on graph theory parameters, and we need these parameters for all systems of interest.

One of the regression approaches applied and reported was using principal component regression (PCR). In PCR, the training data is converted to a data matrix with orthogonalized axes, which are based on capturing the maximum amount of information in fewer dimensions. In this case, as we were predicting for a single property, The relationships discovered in the training data can be applied to a test dataset based on a projection of the data onto a high-dimensional hyperplane within the orthogonalized axis-system. Typical linear regression models do not properly account for the co-linearity between the descriptors, and as a result the isolated impact of each descriptor on the property cannot be accurately known. However, by projecting the data onto a high-dimensional space defined by axes which are comprised of a linear combination of the composite descriptors and also orthogonalized, the impact of the descriptor on the property can be identified independent of all other descriptors.

PCR finds the maximum variance in the predictor variables ($X$) and finds the correlation factors between $X$ and the predicted variables ($Y$) that have maximum variance. The scores of $X$, $t_a$ ($a=1, 2, \ldots, A$=the number of principal components) are calculated as linear combinations of the original variables with the weights $w^*_{ka}$. The multidimensional space of $X$ is reduced to the A-dimensional hyper plane. Since the scores are good predictors of $Y$, the correlation of $Y$ is formed on this hyper plane. The loadings of $X$ ($P$) represent the orientation of each of the components of the hyper plane. Following this, an accurate and high-throughput equation linking the input parameters and the properties are derived.

S2. Descriptor Space

The following (Table S1) provides a list of the input descriptors used in the initial analysis, and provides the list of descriptors from Figure 3. These descriptors have been utilized in our prior work. Of note, these descriptors describe the single element systems, with the values corresponding with the single element systems and in their ground state structures. The scaling of these descriptors to account for multicomponent systems was discussed in the main document. In this case, we limited our search space to $LiMe_2O_4$ and therefore we only scale based on the ratio of elements comprising Me (ie. the B-site in the spinel structure). Future reports will expand to account changing the overall stoichiometry, as well as the A-site and X-site chemistries.

*Table S1. The descriptors used in this analysis. In this paper, we defined descriptors for multi-component systems based solely on the descriptors of the individual elements. In this way, there is no limitation in the 'virtual' design space for which we can apply our model. The uncertainty in models is based solely on the amount of available information in the training or test data, but as this descriptor data is available for nearly all elements, the model can be applied to a nearly infinite chemical space.*

| Interatomic distance | Pauling electronegativity (EN) | Heat of Vaporization |
|---|---|---|
|  |  |  |
| Valence electron number | Martynov-Batsanov EN | Heat Capacity |
| Pseudopotential core radii sum | Melting Point | Melting Point |
| Covalent Radius | Boiling Point | Boiling Point |
| Atomic Radius | Bulk Modulus | Modulus of Elasticity |
| Atomic Weight | Shear Modulus | Electrical Conductivity |
| Molar Volume | Work Function | Thermal Conductivity |
| Density @ 293 K | Specific Heat | Coeff. Thermal Expansion |
| First Ionization Potential | Heat of Fusion |  |

S2. Training Data

In order to minimize the number of factors to capture in the initial model, the training data followed the following constraints. The spinel structure was fixed to $LiMe_2O_4$, with Me allowed to host up to three elements in various ratios and the Me elements allowed defined in Figure 1. The test temperature was confined to room temperature, with the processing temperature between 600 and 800 degrees, and the measurement value used when possible after the first cycle. Future reports on this work will remove these various constraints. The training data reported in Figure 9 are shown in Table S2. Note, the data reported here is for the data reported under these conditions. Additionally, when there are multiple reported values, the value used is the average of all the entries. This data was normalized and then parameterized in order to have a representation of the data applicable for both the testing systems and the new systems which have not been reported previously.

*Table S2. Systems used in training the model and reported in Figure 9. These follow constraints of same stoichiometries, containing same amounts of Li and O, and measurements performed under similar processing and test conditions. From this small data set, we are able to generate large datasets. Future testing of the proposed alloy systems from this paper will provide a further validation, as well as feeding back into the analysis to increase the governing physics and reduce the model uncertainty.*

|  | Average Capacity (mAh/g) |
|---|---|
| LiTi2O4 | 162 |
| LiMnTiO4 | 151 |
| LiMn2O4 | 148 |
| LiCo2O4 | 147 |
| LiCoMnO4 | 146 |
| LiFe2O4 | 142 |
| LiTiVO4 | 141 |
| LiFeNiO4 | 132 |
| LiFeTiO4 | 131 |
| LiCoNiO4 | 129 |
| LiNi0.5Mn1.5O4 | 122 |
| LiNi2O4 | 117 |
| LiCu2O4 | 115 |
| LiCu0.5Mn1.5O4 | 105 |

S3. Rough Set Theory

Rough set theory is an approach to define confidence intervals in material classifications. For this, we first define the known data into three different categories: high capacity, medium capacity and low capacity. The next step is to then find a projection of the data which captures classifications of the data. That is, groupings of the different classes allows us to define target regions. Traditionally, the classification is done by plotting two properties or parameters versus each other. However, in this work we find a projection through the use of PCA (Figure S1), which uses PC1 versus PC3. In this, we find a region which contains only high capacity systems (red circles). The systems we predicted as most promising (open triangles) which fall in this region we then define as having higher confidence in our prediction as a high capacity system. We then define a medium confidence area which contains all high capacity data. Interestingly, this region contains some medium capacity systems (blue circles) but no low capacity systems (green diamonds). This approach then allows us to quickly provide a confidence ranking of our predictions (as shown in Table 2), with each of these open triangles corresponding to a row in the table. This step is important given the level of uncertainty associated with relatively small training data.

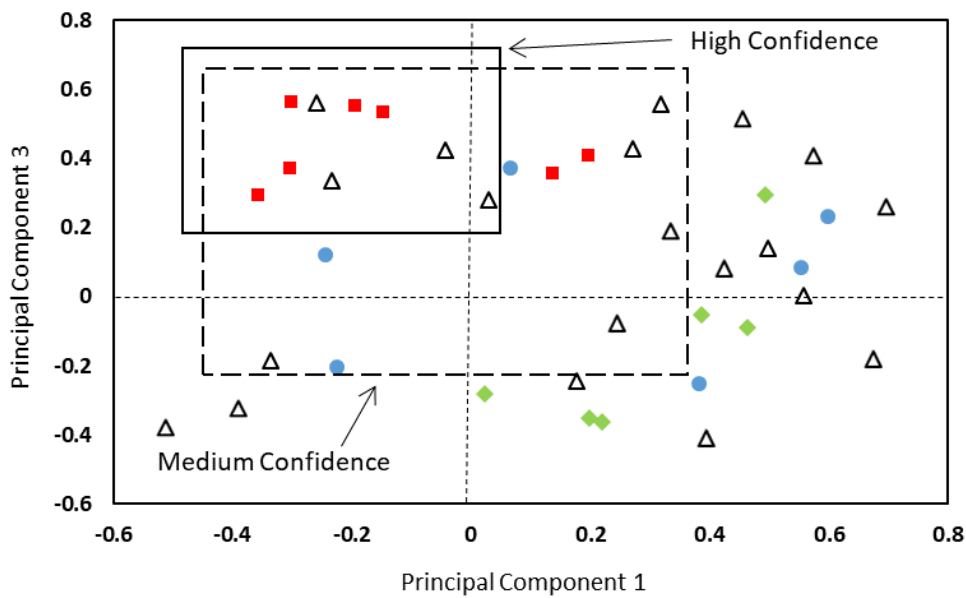

*Figure S1. Result from RST, providing confidence intervals of predictions. While RST typically operates on two descriptors, we have expanded here to defining parameters through principal components. The red squares are those materials which have high capacity, the blue circles have medium capacity, the green diamonds have low capacity, and the open triangles are the compounds we predicted Table S2. The positioning of these compounds in this plot were used to define the confidence levels in Table 2.*

Supplementary References

1. J. Tenenbaum, V. de Silva, J. Langford, A Global Geometric Framework for Nonlinear Dimensionality Reduction, Science, 290, 2319 (2000).
2. A. Mullis, S. Broderick, A. Binnebose, N. Peroutka-Bigus, B. Bellaire, K. Rajan, B. Narasimhan. "Data Analytics Approach for Rational Design of Nanomedicines with Programmable Drug Release." Molecular Pharmaceutics, 16, 1917-1928 (2019)
3. X. Zhen, T. Zhang, S. Broderick, K. Rajan. "Correlative Analysis of Metal Organic Framework Structures through Manifold Learning of Hirshfeld Surfaces." Molecular Systems Design & Engineering. 3, 826-838 (2018)
4. S. Srinivasan, S.R. Broderick, R. Zhang, A. Mishra, S.B. Sinnott, S.K. Saxena, J.M. LeBeau, K. Rajan. "Mapping Chemical Selection Pathways for Designing Multicomponent Alloys: An Informatics Framework for Materials Design." Scientific Reports. 5, 17960 (2015)
5. A. Wagaman, E. Levina, E. Discovering sparse covariance structures with the isomap. *Journal of Computational and Graphical Statistics 18*, 551-572 (2009).

6. W.J. Cukierski, D.J. Foran, D.J. In *Using betweenness centrality to identify manifold shortcuts*, Data Mining Workshops, 2008. ICDMW'08. IEEE International Conference on, 2008; IEEE: 2008; pp. 949-958.
7. M. Balasubramanian, E.L. Schwartz, The Isomap Algorithm and Topological Stability, *Science,* 295, 7 (2002).
8. L. Ericksson, E. Johansson, N. Kettaneh-Wold, S. Wold. Multi- and megavariate data analysis: principles, applications, Umea, SwedenUmetricsAb (2001).
9. S. Broderick, K. Rajan. "Informatics Derived Materials Databases for Multifunctional Properties." Science and Technology of Advanced Materials. 16, 013501 (2015)
10. P.V. Balachandran, S.R. Broderick, K. Rajan. "Identifying the "Inorganic Gene" for High Temperature Piezoelectric Perovskites through Statistical Learning." Proceedings of the Royal Society A. 467, 2271 (2011)